\documentclass[conference]{IEEEtran}
\IEEEoverridecommandlockouts
\usepackage{cite}
\usepackage{amsmath,amssymb,amsfonts}
\usepackage{algorithmic}
\usepackage{graphicx}
\usepackage{textcomp}
\usepackage{xcolor}
\usepackage{support-caption}
\usepackage{subcaption}
\usepackage{setspace} 
\usepackage{xspace}
\usepackage{multirow}
\usepackage{booktabs} % Tables
\usepackage[super]{nth}
\usepackage[bookmarks=false]{hyperref}
\AtBeginDocument{\NoHyper}

\usepackage{flushend}

\newcommand{\fwname}{SPINE}

\newcommand{\captionbelowmargin}{\vspace{-6mm}}

\newcommand{\titlemargin}{\vspace{-5mm}}

\ifx\final\undefined

\else

\fi

\def\BibTeX{{\rm B\kern-.05em{\sc i\kern-.025em b}\kern-.08em
    T\kern-.1667em\lower.7ex\hbox{E}\kern-.125emX}}
\begin{document}

\title{\fwname: A Fault Injection Profiler for Quantized Neural Networks under Accumulated Faults
\vspace{-3mm}

\ifx\blindreview\undefined
\thanks{%
%This work was partially funded by
%the Office of Naval Research Global (ONR) (N629092212014),
%Coordenação de Aperfeiçoamento de Pessoal de Nível Superior - Brasil (CAPES) - Finance Code 001,
%CNPq,
%FAPERGS Techfuturo - 23/2551-0002199-4, 
% PADIS through HT Micron Semicondutores,
%and IoT na Borda, supported by CEDRA, with financial resources from the PPI IoT/Manufatura 4.0 / PPI HardwareBR of the MCTI grant number 056/2023, signed with EMBRAPII.
}
\else
\thanks{This study was financed in part by agencies.}
\fi
}

\ifx\blindreview\undefined

\author{
\IEEEauthorblockN{
Nathan Guimarães\IEEEauthorrefmark{1}\IEEEauthorrefmark{2}, Ian Kersz\IEEEauthorrefmark{1}\IEEEauthorrefmark{2}, Leonardo R. Gobatto\IEEEauthorrefmark{1}\IEEEauthorrefmark{2}, Fabio Benevenuti\IEEEauthorrefmark{1}, Michael G. Jordan\IEEEauthorrefmark{1}\IEEEauthorrefmark{2}, \\
Antonio Carlos S. Beck\IEEEauthorrefmark{1}\IEEEauthorrefmark{2},
Fernanda L. Kastensmidt\IEEEauthorrefmark{1}\IEEEauthorrefmark{2}, 
Jose Rodrigo Azambuja\IEEEauthorrefmark{1}\IEEEauthorrefmark{2}
}
\IEEEauthorblockA{
\IEEEauthorrefmark{1}\textit{Federal University of Rio Grande do Sul (UFRGS) - Institute of Informatics - PPGC - PGMICRO}, Porto Alegre, Brazil\\
\{leonardo.gobatto,  nathan.guimaraes, fabio.benevenuti, caco, fglima, jose.azambuja\}@inf.ufrgs.br\\
}
\IEEEauthorblockA{\IEEEauthorrefmark{2}\textit{Center for Embedded Devices and Research in Digital Agriculture (CEDRA)}, São Leopoldo, Brazil}
\titlemargin % !!DIMINUI O ESPAÇAMENTO ENTRE AUTORES E TEXTO!!
}

\else

\author{
   \IEEEauthorblockN{Omitted for blind review.}
   \vspace{-5mm}
   
}

\fi

\maketitle
\vspace{-5mm}

\begin{abstract}
Deploying deep neural networks at the edge demands efficient inference under strict cost and power constraints. Quantized neural networks address these demands by replacing floating-point parameters with low-precision integers, yet their weights remain continuously exposed to radiation-induced bit-flips during inference. Fault Injection can be used to simulate those environments, but existing studies fail to characterize how accumulated upsets translate into mispredictions under realistic memory layouts. This paper presents a GDB-driven profiling framework that injects cumulative weight bit-flips directly onto the target binary of edge CPUs, generating per-layer fault profiles without requiring model retraining or code modification. Evaluated across multiple topologies, quantization efforts, and memory layouts, the results indicate how selective hardening strategies should be applied to effectively protect neural networks.
\end{abstract}


\vspace{-2mm}

\section{Introduction}
\label{sec:intro}

Deep Neural Networks (DNNs) are increasingly run at the edge, closer to where data is produced~\cite{edge_ai}, such as on a satellite that classifies its captured images on board~\cite{edge_space2}.
This paradigm is reinforced by the \emph{new space} trend, 
where the use of commercial off-the-shelf components is replacing 
traditional radhard circuits to meet the stringent cost, 
time-to-market, and performance demands of modern small-satellite 
missions~\cite{cots_newspace,newspace}. 
Inference on these platforms may run on general-purpose processors rather than on dedicated accelerators, driven by cost, flexibility, and power constraints~\cite{embedded_ai, edge_mcu}.
To reduce the computational and memory costs of running DNNs within these strict budgets, a popular approach is to use quantization, which replaces floating-points with low-precision integers~\cite{quantization}.

In Quantized Neural Networks (QNNs), the weights dominate the memory footprint during inference and are continuously read from RAM as the model runs on each input.
Because memories in these platforms are subject to transient bit-flips caused by radiation, electrical disturbances, and other sources~\cite{radiation_effects_cots}, every bit of every parameter is repeatedly exposed to these faults.
Due to the probabilistic nature of these networks, some bit-flips remain inconsequential, which allows the QNN to match the fault-free reference even as upsets accumulate. In contrast, others can trigger mispredictions, degrading the model's accuracy~\cite{fault_propagation}.
The severity of these errors depends on their location within the network, the weight-encoding scheme, and whether the upsets occur in isolation or in specific combinations.

\begin{figure}[t]
    \centering
    \includegraphics[width=0.9\columnwidth]{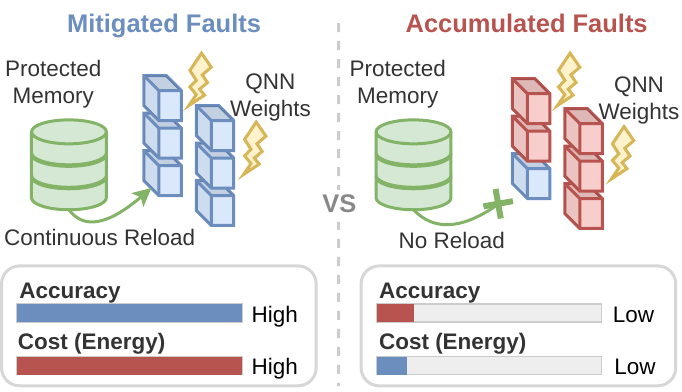}
    \caption{\centering Extremes of the protection/accuracy trade-off.}
    \label{fig:intro}
    \captionbelowmargin
\end{figure}

Fig.~\ref{fig:intro} illustrates the extremes of protecting QNNs against these upsets.
On the left-hand side, continuous weight reload erases data corruption and increases fault tolerance while imposing recurring energy and latency penalties that negate the efficiency of edge deployment.
On the right-hand side, a single weight load without reloads leads to fault accumulation and eventually mispredictions (system failure).
Intermediate strategies, such as selectively reloading highly sensitive weights or layers, require characterizing how individual and accumulated upsets translate into output errors on the target device.
However, existing fault-injection studies fail to provide this characterization, as they largely evaluate faults in isolation and overlook the actual memory layout of compressed weights.

This paper presents \textbf{S}ensitivity \textbf{P}rofiler for \textbf{In}ference on the \textbf{E}dge (\textbf{\fwname}), a framework that profiles QNN sensitivity to weight faults on edge CPUs to generate a per-layer Fault Injection Profile.
Our main contributions are to formulate a weight-sensitivity characterization as a cumulative, misprediction-bounded fault injection problem and to evaluate it across representative QNN topologies and datasets, revealing how layer depth, quantization precision, and memory layout shape the model's vulnerability to accumulated upsets.

\section{Related Work}
The reliability of DNNs under hardware faults has been extensively studied, particularly through Fault Injection (FI) methodologies that aim to quantify model resilience. Previous work, such as Ares~\cite{ares}, introduced a systematic framework for evaluating the impact of transient faults on DNN inference. This framework demonstrated that DNN resilience is highly non-uniform, varying significantly across layers, fault types, and injection locations. This observation motivates selective protection strategies, as uniformly protecting all parameters leads to unnecessary overhead. At a finer granularity, parameter-level vulnerability was investigated, and techniques were proposed to approximate parameter sensitivity with reduced FI effort, highlighting that vulnerability is strongly correlated with layer position and parameter significance~\cite{vulnerability_parameters}.

The fault model most pertinent to our target context is the Single Event Upset (SEU), a transient bit-flip induced by ionizing radiation. SEU fault models for DNNs were formalized and reported non-uniform layer-wise robustness, providing the sensitivity definition and modeling approach that a per-layer characterization framework should adopt~\cite{seu_dnn}. In Aspis~\cite{aspis}, selective protection based on per-layer criticality uses a Taylor-based proxy to identify critical weights and apply targeted hardening.
%Selective protection, guided by per-layer criticality, has since been operationalized in works such as Aspis~\cite{aspis}, which uses a Taylor-based proxy to identify critical weights and applies targeted hardening.
In our hardware context, NeuralFuse~\cite{neuralfuse} targets accuracy degradation caused by SRAM bit-flips at reduced supply voltage in memory-constrained inference and proposes a lightweight add-on module to recover accuracy under such conditions.

However, all of these studies share two key limitations with respect to our work: they evaluate faults in isolation, without modeling cumulative upsets over time; and they largely target floating-point models on GPUs or dedicated accelerators, disregarding the quantized weight representation and actual binary memory layout that characterize QNN inference on edge CPUs. Our framework addresses both gaps by formulating sensitivity characterization as a cumulative, misprediction-bounded injection problem and operating directly on the target binary via GDB.

\section{Framework}
\label{sec:framework}
\begin{figure}[t]
    \centering
    \includegraphics[width=\columnwidth]{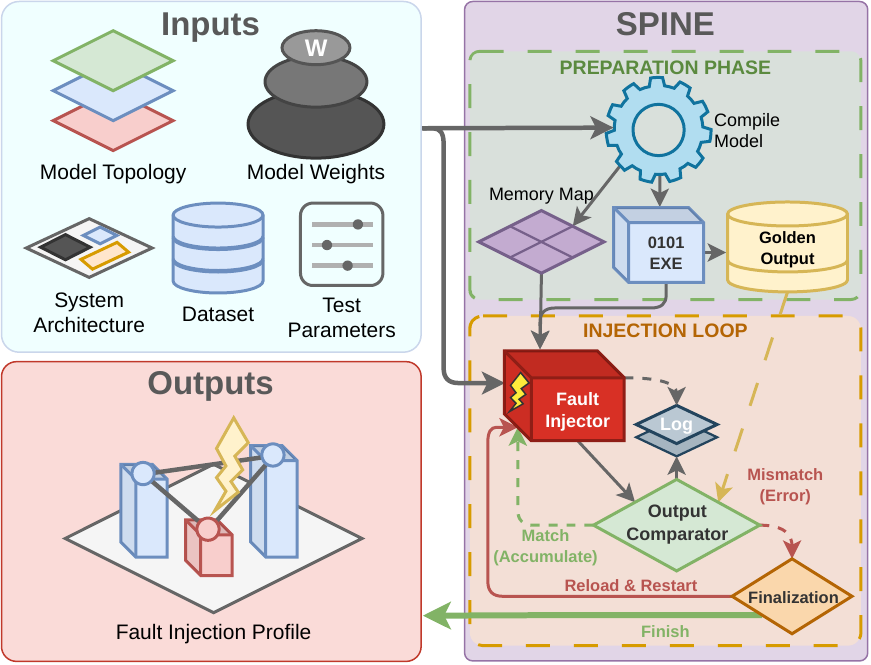}
    \caption{\centering Operational workflow and architecture of \fwname.}
    \label{fig:arch}
    \captionbelowmargin
\end{figure}

%\subsection{Inputs}
%\label{subsec:inputs}

\fwname, shown in Fig.~\ref{fig:arch}, takes five inputs that describe the case under study:
the \emph{Model Topology}, which specifies the network's layer structure;
the \emph{Model Weights}, which provide its parameters;
the \emph{System Architecture}, which identifies the target CPU or microcontroller;
the \emph{Dataset}, which supplies the samples used to exercise the model;
and the \emph{Test Parameters}, which set the stopping criteria and other campaign options.

%\subsection{Framework}
%\label{subsec:framework}

From the provided inputs, \fwname~operates in two sequential phases: a \emph{Preparation Phase}, which builds the artifacts needed for injection, and an \emph{Injection Loop}, which systematically perturbs the model weights and measures the impact on its predictions.
To keep the exploration space tractable, the framework adopts a layer-isolation strategy: rather than injecting faults across all layers simultaneously, a complete injection campaign is conducted for each layer independently, leaving the other layers untouched.
This way, the fault tolerance of each layer can be characterized in isolation before aggregation.

\textbf{Preparation Phase}.
%\label{subsubsec:preparation}
The \emph{Compile Model} step produces the two artifacts needed during injection.
The first is the executable binary of the inference program, together with its \emph{Memory Map}, which records the address range occupied by the quantized weights on the target and indexes every address back to the originating layer.
The second is the \emph{Golden Output}, obtained by executing the fault-free binary on the dataset and collecting its outputs as the reference against which later faulty executions will be compared.
Once these artifacts are ready, \fwname~enters the injection loop.

\textbf{Injection Loop}.
%\label{subsubsec:loop}
Iterates over each layer in isolation and, within each layer campaign, is driven by three mechanisms.

The \emph{Fault Injector} starts each iteration by applying one new bit-flip to the weight region of the binary running under debugger control.
Built on top of an open fault injector~\cite{leo_gdb}, it attaches to the target processor via the GNU Debugger (GDB) and modifies the weights in place without changing the inference code or the compiled model.
A random bit position is read from the address range exported by the \emph{Memory Map} for the layer under campaign, and its current value is flipped through a single write at the target address.
Because the \emph{Memory Map} associates every address with the layer it belongs to, each flip is recorded alongside the layer and bit position it affected, producing the per-flip trace that is put inside a \emph{Log}.
With the new flip now in place on top of those already accumulated within the current layer's campaign, control passes to the \emph{Output Comparator}.

The \emph{Output Comparator} runs the inference binary on every sample in the dataset and checks each output against the corresponding \emph{Golden Output} entry.
Any output that disagrees with its reference is written to a \emph{Log} with the current flip combination and the affected bit positions, while the pass continues across all remaining samples.
A pass with no logged mismatch keeps the accumulated faults in place and lets the loop proceed to the next injection with one more fault on top; a pass with at least one mismatch is treated as a misprediction event, preserving the logged combinations and handing control to the \emph{Finalization} step.

The \emph{Finalization} is reached after any pass that produced at least one logged misprediction, which we refer to as a Silent Data Corruption (SDC). It checks whether the \emph{Target} defined in the \emph{Test Parameters} has been reached, such as the configured number of SDCs collected.
If it has not, the weights are reloaded from the original quantized parameters, and the injection resumes from a clean state, so that each accumulation starts without carryover from previous ones.
Once the target for the current layer is met, the loop moves to the next layer and repeats the campaign from scratch.

%\subsection{Outputs}
%\label{subsec:outputs}

After all campaigns are complete, the collected logs are aggregated into the \emph{Fault Injection Profile}, reporting per layer how many accumulated flips the network tolerated before mispredicting and which bit positions were most often involved.

\section{Evaluation}
\label{sec:methodology}

We evaluate \fwname~on an Arm Cortex-M3 core embedded in the Cypress PSoC 5LP SoC, mounted on a CY8CKIT-059 development board~\cite{cypress}.
The core implements the Armv7-M profile, has no hardware FPU, and addresses memory at byte granularity, making integer-arithmetic inference on byte-aligned weights the natural strategy for this class of device.
All inference binaries are compiled and executed natively on the board, so that the memory layout and weight addressing seen by \fwname~reflect real deployment conditions.

\begin{table}[!t]
\caption{Structural details of the evaluated topologies.}
\label{tab:topologies}
\renewcommand{\arraystretch}{1.1}
\centering
\resizebox{\columnwidth}{!}{%
\begin{tabular}{lcccccc}
\hline
Topo. & Quant. & \# Layers & \# Weights & \# MACCs (k) & Acc. (\%) \\
\hline \hline
\multirow{2}{*}{$T_1$} & 4 & \multirow{2}{*}{2} &  \multirow{2}{*}{840} & \multirow{2}{*}{215} & 98.78 \\
&8&&&& 98.99 \\ \hline
\multirow{2}{*}{$T_2$} & 4 & \multirow{2}{*}{4} &  \multirow{2}{*}{3744} & \multirow{2}{*}{243}  & 97.99 \\
&8&&&& 98.76  \\ \hline
\multirow{2}{*}{$T_3$} & 4 & \multirow{2}{*}{3} & \multirow{2}{*}{3744} & \multirow{2}{*}{350} & 99.00 \\
&8&&&& 99.27 \\ 
\hline
\end{tabular}%
}
\end{table}

%\subsection{Models and Configurations}
%\label{subsec:configs}

Three neural network topologies, referred to as $T_1$, $T_2$, and $T_3$, are evaluated on the SAT-6 aerial image dataset~\cite{sat6}, a benchmark of satellite imagery labeled across six land-cover categories, representative of on-orbit inference workloads. 
Table~\ref{tab:topologies} lists the structural details of each combination of topology and quantization level, including its number of layers, weights, Multiply-Accumulate (MACCs) operations, and accuracy.
Each is trained on SAT-6 and then quantized at two weight precisions, yielding six baseline models. 
The original SAT-6 images have dimensions of $28 \times 28$ pixels with four channels (red, green, blue, and near-infrared). However, for our experiments, we apply zero-padding to obtain inputs of size $32 \times 32$ while preserving the four-channel structure.
The 8-bit variants store one weight per byte, aligning directly with the target's addressing.
The 4-bit variants halve the precision, but since the core has no native sub-byte type, their weights can be arranged in memory in two different ways:

\textbf{Unpacked (\texttt{U}):}
4-bit weights occupy their own byte through sign extension, preserving the one-weight-per-byte alignment and enabling direct memory access at the cost of doubling the memory footprint relative to a compact 4-bit layout.

\textbf{Packed (\texttt{P}):}
Two weights share each byte, fully exploiting the 4-bit precision to halve memory usage, at the cost of masking and shifting on every weight access during inference.

The 4-bit topologies are compiled in both representations, contributing six configurations that join the three 8-bit ones for a total of \emph{nine} subjected to the injection campaigns. Each of these configurations is then compiled into a standalone Cortex-M3 binary and submitted to \fwname, which runs the per-layer campaign described in Table~\ref{sec:framework}.
All campaigns share the same \emph{Test Parameters}, which are meant to inject enough faults until each network layer reaches 100 misclassifications.
% TODO: replace [X], [Y], [Z] with the actual values used in the experiments.
%Keeping these fixed ensures that the resulting Fault Injection Profiles differ only along the dimensions that define the configurations: topology, precision, and, for the 4-bit variants, memory layout.

A statistical fault injection methodology~\cite{statistical_fi} defines the minimum number of faults required to achieve a desired error margin in a DNN fault injection campaign. Since our campaign was conducted independently of this formulation, we rearrange it to assess the statistical representativeness of our sample:

\begin{equation}
    e = t \cdot \sqrt{\frac{p\,(1-p)\,(N - n)}{n\,(N-1)}}
    \label{eq:error_margin}
\end{equation}

\noindent in which $N$ is the total fault population, defined as the total number of bits across each layer of the network, $n$ is the number of injected faults, $t$ is the constant associated with the desired confidence level (CL), and $p$ is the probability of a fault becoming a critical failure. Since our campaign employs an accumulated-fault model, an observed failure may result from the combination of injected faults rather than a single fault, making it impossible to estimate $p$ empirically. We therefore adopt the conservative worst-case assumption $p = 0.5$, maximizing $p(1-p)$ and, consequently, the error margin. This equation will be applied to evaluate the representativeness of our fault injection campaign at a CL of 99\%.

\section{Results}

The following analysis evaluates the fault injection campaigns across the tested configurations, focusing on per-layer SDC rates, bit-level vulnerability, the impact of memory layout, and cumulative error tolerance. Throughout the graphs, the quantization levels of 8-bits (Q8), 4-bits unpacked (Q4U), and 4-bits packed (Q4P) are represented as bars with the colors green, blue, and orange, respectively. $T_1$, $T_2$, and $T_3$, are separated from each other by the background color in the graphs, while their layers (L) are numerated from 1 to 4.

\begin{figure}[t]
    \centering
    \includegraphics[width=\columnwidth]{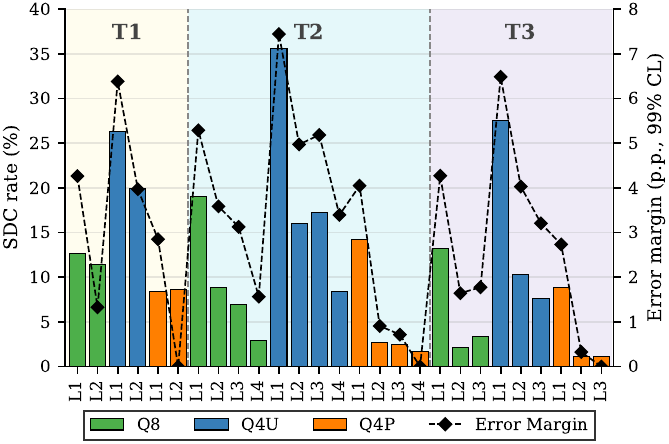}
    \caption{\centering Per-layer SDC rate and error margin across configurations.}
    \label{fig:sdc_rate}
    \captionbelowmargin
\end{figure}

Fig.~\ref{fig:sdc_rate} plots the per-flip SDC rate alongside the statistical error margin for each layer across all experimental configurations. The data reveals a strong stratification of vulnerability by depth, with the input-proximal (first) layers exhibiting the highest susceptibility to missclassifications. The error margin scales inversely with the number of injections, approaching zero for fully sampled layers, which confirms the statistical representativeness of this depth-wise vulnerability trend. 

\begin{figure}[t]
    \centering
    \includegraphics[width=\columnwidth]{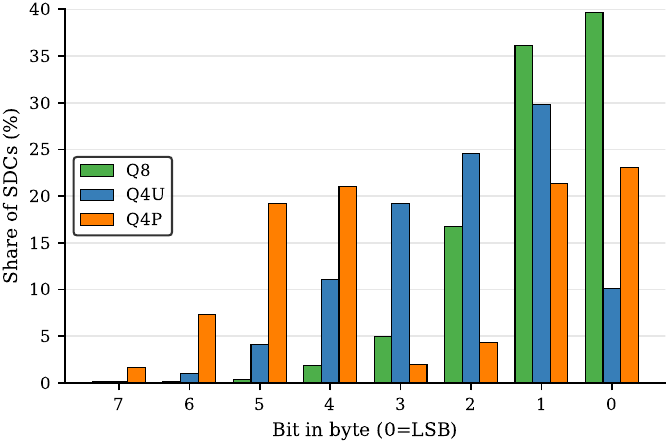}\hfill
    \caption{\centering Distribution of SDC events over the eight bit-positions within each byte, split by quantization regime.}
    \label{fig:bit_in_byte}
    %\captionbelowmargin
\end{figure}

Fig.~\ref{fig:bit_in_byte} decomposes the share of SDC events across the eight bit-positions within a byte for each quantization regime, in average for all layers and topologies. The distribution is asymmetric, indicating that SDCs are overwhelmingly driven by faults in low-order data bits rather than high-magnitude bits. Notably, the sign bit in Q8 models and the unused upper nibble in Q4U models rarely produce SDCs; instead, they act as natural fault sinks that safely trigger detectable application-level crashes when corrupted.

\begin{figure}[t]
    \centering
    \includegraphics[width=\columnwidth]{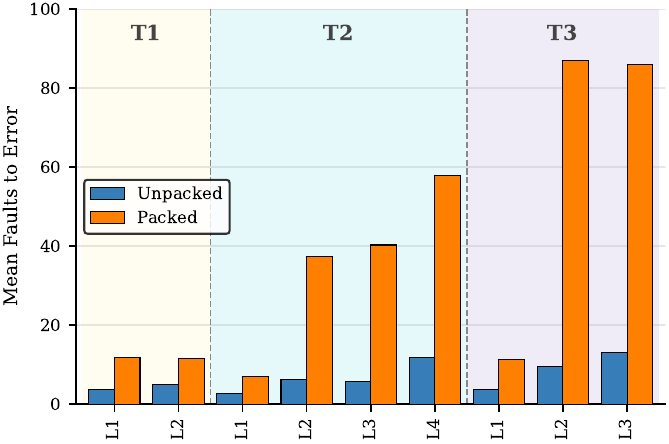}
    \caption{\centering Comparison of unpacked and packed 4-bit memory layouts across matched topology layers.}
    \label{fig:layout}
    \captionbelowmargin
\end{figure}

Fig.~\ref{fig:layout} compares the unpacked and packed 4-bit memory layouts across matched topology layers, evaluating them on mean faults to SDC. Contrary to initial architectural assumptions, memory packing severely degrades model reliability. Unpacked models systematically exhibit lower SDC rates and higher tolerance to faults because their zero-padded bits safely absorb radiation events. By eliminating this inherent redundancy, packing ensures that nearly every fault affects critical weight data, accelerating the system's failure.

\begin{figure}[t]
    \centering
    \includegraphics[width=\columnwidth]{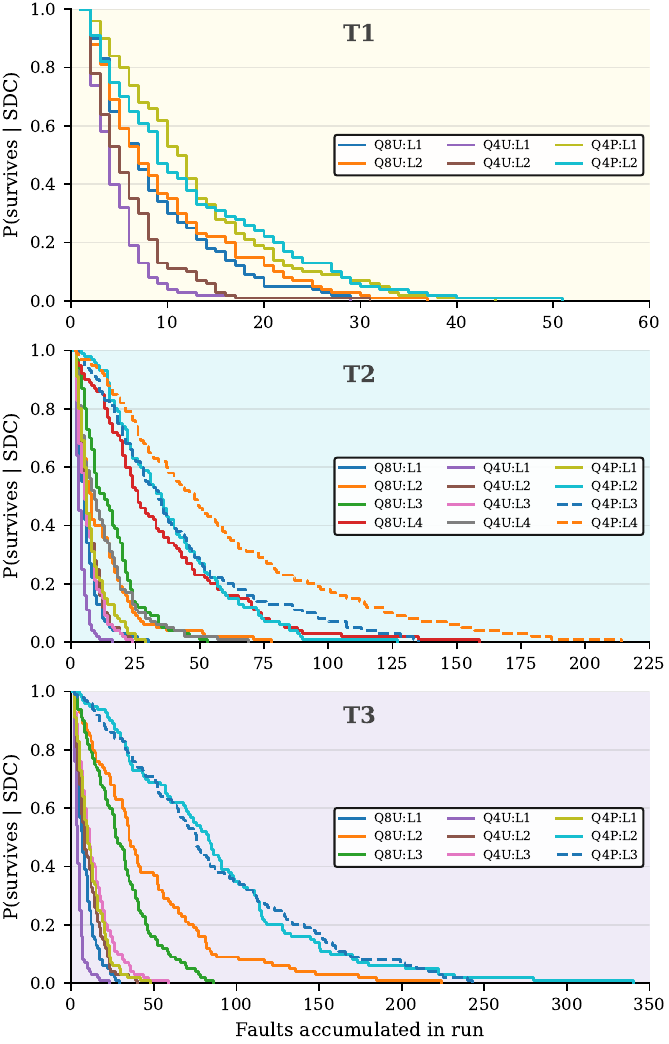}
    \caption{\centering Survival probability curves per topology.}
    \label{fig:survival}
    \captionbelowmargin
\end{figure}

Fig.~\ref{fig:survival} traces the survival probability curves for individual runs, illustrating the likelihood that a layer avoids an SDC as sequential bit-flips accumulate. The varying slopes confirm that resilience must be profiled per-layer rather than globally. Packed configurations exhibit steep, brittle drop-offs, whereas unpacked configurations and deeper topological layers feature long survival tails, indicating a robust capacity to absorb multiple accumulated faults before a critical failure.

\section{Conclusion}
This paper presented SPINE, a framework to profile quantized neural network sensitivity to accumulated weight bit-flips on edge microcontrollers, capturing fault injection profiles and survival probabilities across multiple topologies and quantization schemes. Results show that memory packing degrades resilience by eliminating protective padding bits, low-order bits drive most silent data corruptions, and input-proximal layers overwhelmingly determine the network's failure point, implying selective protection must prioritize shallow layers in unpacked layouts. Future work will leverage these profiles to implement an energy-efficient selective weight reloading system as the intermediate strategy between the two extremes introduced in the introduction and also to measure its gains in simulation and under real radiation doses.

\section*{Acknowledgment}
This work was partially funded by Coordenação de Aperfeiçoamento de Pessoal de Nível Superior - Brasil (CAPES) - Finance Code 001, CNPq, FAPERGS Techfuturo - 23/2551-0002199-4, and by the Center for Embedded Devices and Research in Digital Agriculture (CEDRA) of SENAI-RS, with financial resources from the PPI IoT/Manufatura 4.0 / PPI HardwareBR of the MCTI, grant number 056/2023, signed with EMBRAPII.

% \section*{References}

\bibliographystyle{IEEEtran}
\bibliography{references}

\end{document}